\documentclass[twocolumn,aps]{revtex4-1}
\usepackage{graphicx}
\usepackage{amssymb}
\usepackage{amsmath}
\usepackage{subfigure}
\usepackage{graphpap}
\usepackage{dcolumn} 
\usepackage{bm}
\usepackage{color,calc}
\usepackage{longtable}
\usepackage{color}
\usepackage{tikz}

\newcommand\ie{\textit{i.e.}~}

\usepackage{multirow}
\newcommand{\Eq}[1]{Eq.~(\ref{#1})}

\newcommand{\Fig}[1]{Figure \ref{#1}}
\newcommand{\Tab}[1]{Table \ref{#1}}

\newcommand{\erf}{{\rm erf}}
\renewcommand{\c}{{\rm c}}
\newcommand{\D}{\Delta}

\newcommand{\lr}{^{{\rm lr},\mu}}
\newcommand{\sr}{^{{\rm sr},\mu}}
\renewcommand{\b}[1]{\ensuremath{\mathbf{#1}}}
\renewcommand{\d}{\ensuremath{\text{d}}}
\renewcommand{\H}{\ensuremath{\text{H}}}
\newcommand{\RSH}{\ensuremath{\text{RSH}}}
\newcommand{\bra}[1]{\ensuremath{\langle #1 \vert}}
\newcommand{\ket}[1]{\ensuremath{\vert #1  \rangle}}

\begin{document}

\preprint{}
\title{Basis convergence of range-separated density-functional theory}
\author{Odile Franck$^{1,2,3}$}\email{odile.franck@etu.upmc.fr}
\author{Bastien Mussard$^{1,2,3}$}\email{bastien.mussard@upmc.fr}
\author{Eleonora Luppi$^{1,2}$}\email{eleonora.luppi@upmc.fr}
\author{Julien Toulouse$^{1,2}$}\email{julien.toulouse@upmc.fr}
\affiliation{
$^1$Sorbonne Universit\'es, UPMC Univ Paris 06, UMR 7616, Laboratoire de Chimie Th\'eorique, F-75005 Paris, France\\
$^2$CNRS, UMR 7616, Laboratoire de Chimie Th\'eorique, F-75005 Paris, France\\
$^3$Sorbonne Universit\'es, UPMC Univ Paris 06, Institut du Calcul et de la Simulation, F-75005, Paris, France
}

\date{January 30, 2015}
\begin{abstract}

Range-separated density-functional theory is an alternative approach to Kohn-Sham density-functional theory. The strategy of range-separated density-functional theory consists in separating the Coulomb electron-electron interaction into long-range and short-range components, and treating the long-range part by an explicit many-body wave-function method and the short-range part by a density-functional approximation. Among the advantages of using many-body methods for the long-range part of the electron-electron interaction is that they are much less sensitive to the one-electron atomic basis compared to the case of the standard Coulomb interaction. Here, we provide a detailed study of the basis convergence of range-separated density-functional theory. We study the convergence of the partial-wave expansion of the long-range wave function near the electron-electron coalescence. We show that the rate of convergence is exponential with respect to the maximal angular momentum $L$ for the long-range wave function, whereas it is polynomial for the case of the Coulomb interaction. We also study the convergence of the long-range second-order M{\o}ller-Plesset correlation energy of four systems (He, Ne, N$_2$, and H$_2$O) with the cardinal number $X$ of the Dunning basis sets cc-p(C)V$X$Z, and find that the error in the correlation energy is best fitted by an exponential in $X$. This leads us to propose a three-point complete-basis-set extrapolation scheme for range-separated density-functional theory based on an exponential formula.
\end{abstract}

\maketitle

\section{Introduction}

Range-separated density-functional theory (DFT) (see, e.g., Ref.~\onlinecite{TouColSav-PRA-04}) is an attractive approach for improving the accuracy of Kohn-Sham DFT~\cite{HohKoh-PR-64,KohSha-PR-65} applied with usual local or semi-local density-functional approximations. This approach is particularly relevant for the treatment of electronic systems with strong (static) or weak (van der Waals) correlation effects. The strategy of range-separated DFT consists in separating the Coulomb electron-electron interaction into long-range and short-range components, and treating the long-range part by an explicit many-body wave-function method and the short-range part by a density-functional approximation. In particular, for describing systems with van der Waals dispersion interactions, it is appropriate to use methods based on many-body perturbation theory for the long-range part such as second-order perturbation theory~\cite{AngGerSavTou-PRA-05,GerAng-CPL-05b,GerAng-JCP-07,Ang-PRA-08,FroJen-PRA-08,GolLeiManMitWerSto-PCCP-08,JanScu-PCCP-09,FroCimJen-PRA-10,ChaStoWerLei-MP-10,ChaJacAdaStoLei-JCP-10,FroJen-JCP-11,KulSau-CP-12,CorStoJenFro-PRA-13}, coupled-cluster theory~\cite{GolWerSto-PCCP-05,GolWerStoLeiGorSav-CP-06,GolStoThiSch-PRA-07,GolWerSto-CP-08,GolErnMoeSto-JCP-09}, or random-phase approximations~\cite{TouGerJanSavAng-PRL-09,JanHenScu-JCP-09,JanHenScu-JCP-09b,JanScu-JCP-09,ZhuTouSavAng-JCP-10,TouZhuAngSav-PRA-10,PaiJanHenScuGruKre-JCP-10,TouZhuSavJanAng-JCP-11,AngLiuTouJan-JCTC-11,IreHenScu-JCP-11,GouDob-PRB-11,CheMusAngRei-CPL-12,MusSzaAng-JCTC-14}.

Among the advantages of using such many-body methods for the long-range part only of the electron-electron interaction is that they are much less sensitive to the one-electron atomic basis compared to the case of the standard Coulomb interaction. This has been repeatedly observed in calculations using Dunning correlation-consistent basis sets~\cite{Dun-JCP-89} for second-order perturbation theory~\cite{AngGerSavTou-PRA-05,GerAng-JCP-07,JanScu-PCCP-09,KulSau-CP-12,CorStoJenFro-PRA-13}, coupled-cluster theory~\cite{GolWerSto-PCCP-05} and random-phase approximations~\cite{TouGerJanSavAng-PRL-09,JanHenScu-JCP-09,JanScu-JCP-09,TouZhuAngSav-PRA-10,IreHenScu-JCP-11}. The physical reason for this reduced sensitivity to the basis is easy to understand. In the standard Coulomb-interaction case, the many-body wave-function method must describe the short-range part of the correlation hole around the electron-electron coalescence which requires a lot of one-electron basis functions with high angular momentum. In the range-separation case, the many-body method is relieved from describing the short-range part of the correlation hole, which is instead built in the density-functional approximation. The basis set is thus only used to describe a wave function with simply long-range electron-electron correlations (and the one-electron density) which does not require basis functions with very high angular momentum.

In the case of the Coulomb interaction, the rate of convergence of the many-body methods with respect to the size of the basis has been well studied. It has been theoretically shown that, for the ground-state of the helium atom, the partial-wave expansion of the energy calculated by second-order perturbation theory or by full configuration interaction (FCI) converges as $L^{-3}$ where $L$ in the maximal angular momentum of the expansion~\cite{Sch-PR-62,Sch-INC-63,CarSilMet-JCP-79,Hil-JCP-85,God-SJMA-09}. Furthermore, this result has been extended to arbitrary atoms in second-order perturbation theory~\cite{KutMor-JCP-92,GriLud-JPB-02}. This has motivated the proposal of a scheme for extrapolating the correlation energy to the complete-basis-set (CBS) limit based on a $X^{-3}$ power-law dependence of the correlation energy on the cardinal number $X$ of the Dunning hierarchical basis sets~\cite{HelKloKocNog-JCP-97,HALKIER98}. This extrapolation scheme is widely used, together with other more empirical extrapolation schemes~\cite{Fel-JCP-92,Fel-JCP-93,PetKenDun-JCP-93b,PetDun-JPC-95,Mar-CPL-96,Tru-CPL-98,Fel-JCP-13}. In the case of range-separated DFT the rate of convergence of the many-body methods with respect to the size of the basis has never been carefully studied, even though the reduced sensitivity to the basis is one of the most appealing feature of this approach.

In this work, we provide a detailed study of the basis convergence of range-separated DFT. First, we review the theory of range-separated DFT methods (Section~\ref{sec:rsdft}) and we study the convergence of the partial-wave expansion of the long-range wave function near the electron-electron coalescence. We show that the rate of convergence is exponential with respect to the maximal angular momentum $L$ (Section~\ref{sec:pwexpand}). Second, we study the convergence of the long-range second-order M{\o}ller-Plesset (MP2) correlation energy of four systems (He, Ne, N$_2$, and H$_2$O) with the cardinal number $X$ of the Dunning basis sets, and find that the error in the correlation energy is best fitted by an exponential in $X$. This leads us to propose a three-point CBS extrapolation scheme for range-separated DFT based on an exponential formula (Section~\ref{sec:basisconv}).

Hartree atomic units are used throughout this work.

\section{Range-separated density-functional theory}
\label{sec:rsdft}

In range-separated DFT, the exact ground-state energy of an electronic system is expressed as a minimization over multideterminantal wave functions $\Psi$ (see, e.g., Ref.~\onlinecite{TouColSav-PRA-04})
\begin{eqnarray}
E &=& \min_{\Psi} \Bigl\{ \bra{\Psi} \hat{T} + \hat{V}_\text{ne} +\hat{W}_\text{ee}\lr \ket{\Psi} + E_{\H\text{xc}}\sr[n_{\Psi}]\Bigl\},
\label{EminPsi}
\end{eqnarray}
where $\hat{T}$ is the kinetic-energy operator, $\hat{V}_\text{ne}$ is the nuclear--electron interaction operator, $E_{\H\text{xc}}\sr[n_\Psi]$ is the short-range Hartree--exchange--correlation density functional (evaluated at the density of $\Psi$), and $\hat{W}_\text{ee}\lr = (1/2) \iint w_\text{ee}\lr(r_{12}) \hat{n}_2(\b{r}_1,\b{r}_2) \d\b{r}_1 \d\b{r}_2$ is the long-range electron-electron interaction operator written in terms of the pair-density operator $\hat{n}_2(\b{r}_1,\b{r}_2)$. In this work, we define the long-range interaction $w_\text{ee}\lr(r_{12})$ with the error function
\begin{equation}
w_\text{ee}\lr(r_{12}) = \frac{\erf(\mu r_{12})}{r_{12}},
\end{equation}
where $r_{12}$ is the distance between two electrons and $\mu$ (in bohr$^{-1}$) controls the range of the separation, with $r_c = 1/\mu$ acting as a smooth cutoff radius. For $\mu=0$, the long-range interaction vanishes and range-separated DFT reduces to standard Kohn-Sham DFT. In the opposite limit $\mu\to\infty$, the long-range interaction becomes the Coulomb interaction and range-separated DFT reduces to standard wave-function theory. In practical applications, one often uses $\mu \approx 0.5$ bohr$^{-1}$~\cite{GerAng-CPL-05,FroTouJen-JCP-07}.

The minimizing wave function $\Psi\lr$ in Eq.~(\ref{EminPsi}) satisfies the Schr\"odinger-like equation
\begin{eqnarray}
\left( \hat{T} + \hat{W}_\text{ee}\lr + \hat{V}_\text{ne} + \hat{V}_\text{Hxc}\sr[n_{\Psi\lr}] \right) \ket{\Psi\lr} = {\cal E}\lr \ket{\Psi\lr},
\nonumber\\
\label{HPsiEPsi}
\end{eqnarray}
where $\hat{V}_\text{Hxc}\sr$ is the short-range Hartree--exchange--correlation potential operator (obtained by taking the functional derivative of $E_{\H\text{xc}}\sr$), and ${\cal E}\lr$ is the eigenvalue associated with $\Psi\lr$. 

In practice, many-body perturbation theory can be used to solve Eq.~(\ref{HPsiEPsi}). An appropriate reference for perturbation theory is the range-separated hybrid (RSH) approximation~\cite{AngGerSavTou-PRA-05} which is obtained by limiting the search in Eq.~(\ref{EminPsi}) to single-determinant wave functions $\Phi$
\begin{eqnarray}
E_\RSH^\mu &=& \min_{\Phi} \Bigl\{ \bra{\Phi} \hat{T} + \hat{V}_\text{ne} +\hat{W}_\text{ee}\lr \ket{\Phi} + E_{\H\text{xc}}\sr[n_{\Phi}]\Bigl\}.
\label{EminPhi}
\end{eqnarray}
The corresponding minimizing wave function will be denoted by $\Phi^\mu$. The exact ground-state energy is then expressed as
\begin{eqnarray}
E = E_\RSH^\mu + E_\text{c}\lr,
\end{eqnarray}
where $E_\text{c}\lr$ is the long-range correlation energy which is to be approximated by perturbation theory. For example, in the long-range variant of MP2 perturbation theory, the long-range correlation energy is~\cite{AngGerSavTou-PRA-05}
\begin{eqnarray}
E_\text{c,MP2}\lr &=& \bra{\Phi^\mu} \hat{W}_\text{ee}\lr \ket{\Psi^{\text{lr},\mu}_1},
\label{MP2}
\end{eqnarray}
where $\Psi^{\text{lr},\mu}_1$ is the first-order correction to the wave function $\Psi\lr$ (with intermediate normalization). In the basis of RSH spin orbitals $\{\phi_k^\mu\}$, $E_\text{c,MP2}\lr$ takes a standard MP2 form
\begin{eqnarray}
E_\text{c,MP2}\lr = \phantom{xxxxxxxxxxxxxxxxxxxxxxxxxxxxxxx}\nonumber\\
\sum_{i<j}^\text{occ} \sum_{a<b}^{\text{vir}} \frac{\left|\bra{\phi_i^\mu \phi_j^\mu} \hat{w}_\text{ee}\lr\ket{\phi_a^\mu \phi_b^\mu} - \bra{\phi_i^\mu \phi_j^\mu} \hat{w}_\text{ee}\lr\ket{\phi_b^\mu \phi_a^\mu}\right|^2}{\varepsilon_{i}^\mu +\varepsilon_{j}^\mu -\varepsilon_{a}^\mu -\varepsilon_{b}^\mu},
\nonumber\\
\label{MP2orb}
\end{eqnarray}
where $\bra{\phi_i^\mu \phi_j^\mu} \hat{w}_\text{ee}\lr\ket{\phi_a^\mu \phi_b^\mu}$ are the long-range two-electron integrals and $\varepsilon_{k}^\mu$ are the RSH orbital energies. The long-range correlation energy can also be approximated beyond second-order perturbation theory by coupled-cluster~\cite{GolWerSto-PCCP-05} or random-phase~\cite{TouGerJanSavAng-PRL-09,JanHenScu-JCP-09,PaiJanHenScuGruKre-JCP-10,TouZhuSavJanAng-JCP-11,AngLiuTouJan-JCTC-11} approximations. Beyond perturbation theory approaches, Eq.~(\ref{HPsiEPsi}) can be (approximately) solved using configuration interaction~\cite{LeiStoWerSav-CPL-97,PolSavLeiSto-JCP-02,TouColSav-PRA-04} or multiconfigurational self-consistent field~\cite{PedJen-JJJ-XX,FroTouJen-JCP-07,FroReaWahWahJen-JCP-09} methods. Alternatively, it has also been proposed to use density-matrix functional approximations for the long-range part of the calculation~\cite{Per-PRA-10,RohTouPer-PRA-10}.

Since the RSH scheme of Eq.~(\ref{EminPhi}) simply corresponds to a single-determinant hybrid DFT calculation with long-range Hartree-Fock (HF) exchange, it is clear that the energy $E_\RSH^\mu$ has an exponential basis convergence, just as standard HF theory~\cite{HALKIER99}. We will thus focus our study on the basis convergence of the long-range wave function $\Psi\lr$ and the long-range MP2 correlation energy $E_\text{c,MP2}\lr$.

\begin{figure*}
\includegraphics[scale=0.30,angle=-90]{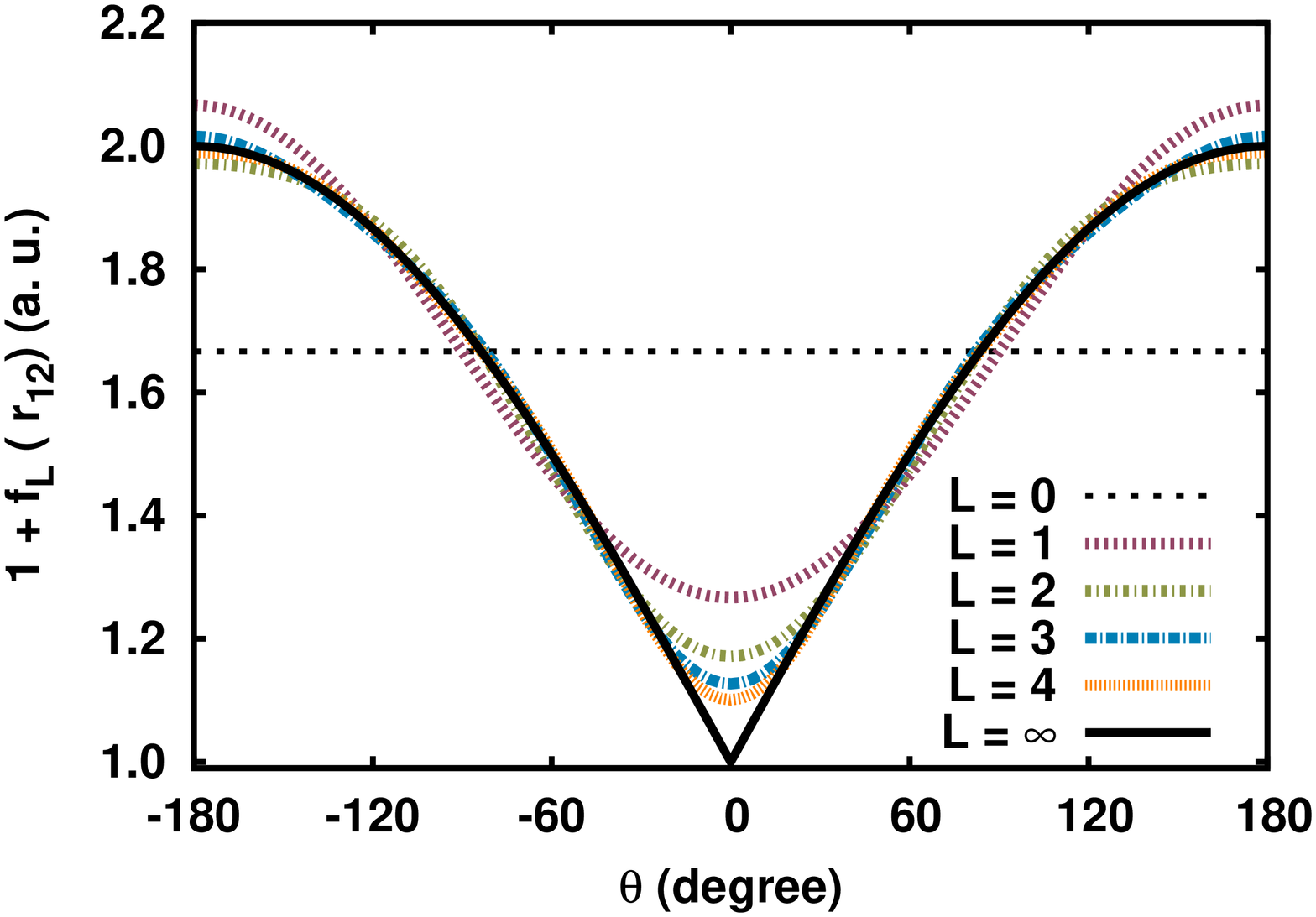}
\includegraphics[scale=0.30,angle=-90]{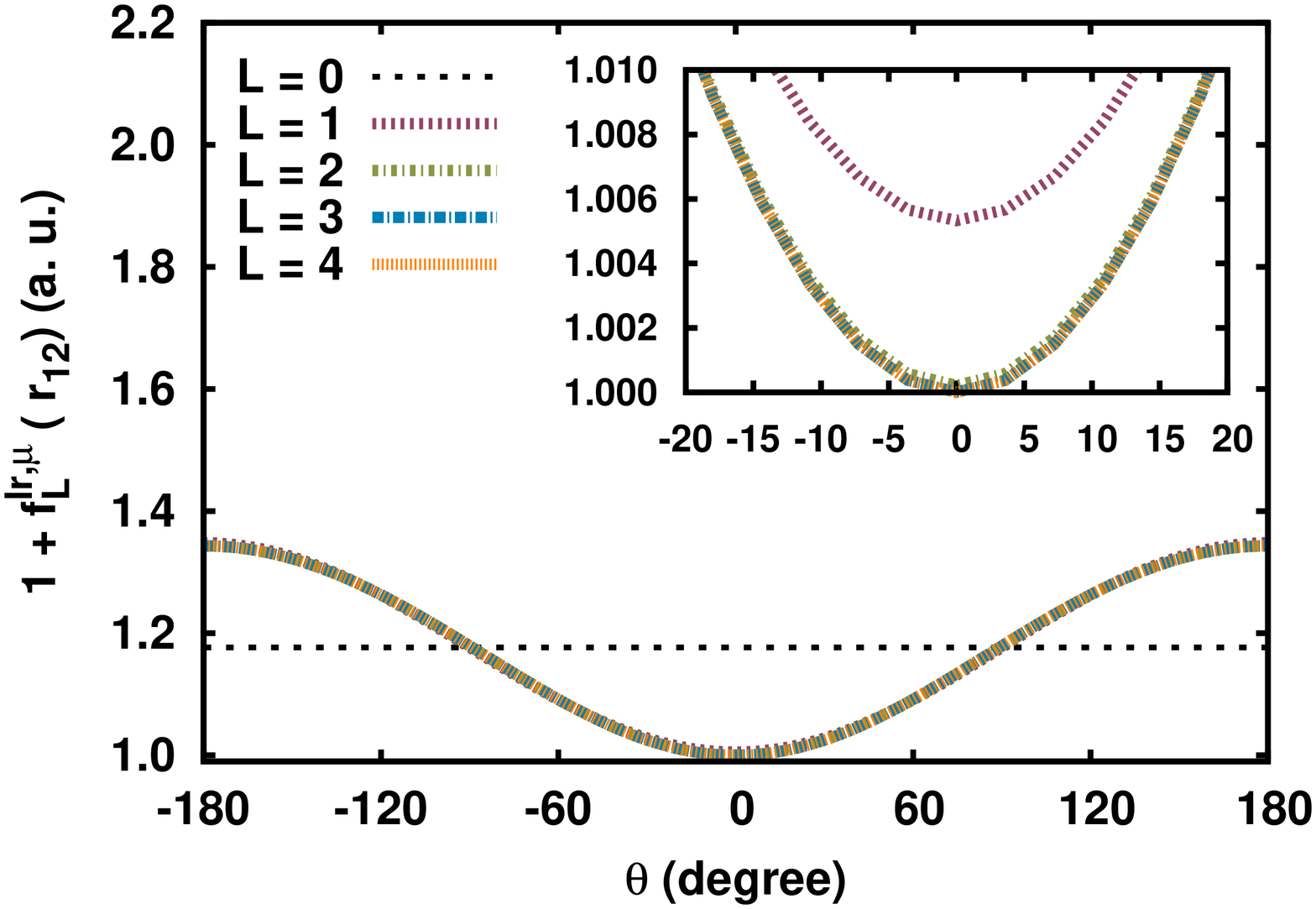}
\caption{
Convergence of the truncated partial-wave expansion $1+f_{L}(r_{12})$ for the Coulomb interaction (left) and $1+f\lr_{L}(r_{12})$ for the long-range interaction using $\mu = 0.5$ bohr$^{-1}$ (right) for different values of the maximal angular momentum $L$. The functions are plotted with respect to the relative angle $\theta$ between the position vectors $\b{r}_1$ and $\b{r}_2$ of the two electrons, using $r_{12}=\sqrt{r_1^2 + r_2^2 -2 r_1 r_2 \cos \theta}$. We have chosen $r_1 = r_2 = 1$ bohr, giving $r_{12}=\sqrt{2 -2 \cos \theta}$. In the insert plot on the right, the curves for $L=2$, $3$, and $4$ are superimposed.}
\label{cusp_plot}
\end{figure*}

\section{Partial-wave expansion of the wave function near electron-electron coalescence}
\label{sec:pwexpand}
In this section, we study the convergence of the partial-wave expansion of the wave function at small interelectronic distances, \ie near the electron-electron coalescence, which for the case of the Coulomb interaction determines the convergence of the correlation energy.
We first briefly review the well-known case of the Coulomb interaction and then consider the case of the long-range interaction.

\subsection{Coulomb interaction}
\label{coulomb_case}
   
For systems with Coulomb electron-electron interaction $w_{\text{ee}}(r_{12})=1/r_{12}$, the electron-electron cusp condition~\cite{Kat-CPAM-57} imposes the wave function to be linear with respect to $r_{12}$ when $r_{12} \rightarrow 0$~\cite{KUTZELNIGG85}
\begin{equation}
\frac{\Psi(r_{12})}{\Psi(0)} =  1 + \frac{1}{2} r_{12} + O(r_{12}^2).
\end{equation}
Here and in the rest of this section, we consider only the dependence of the wave function on $r_{12}$ and we restrict ourselves to the most common case of the two electrons being in a natural-parity singlet state~\cite{KutMor-JCP-92} for which $\Psi(0) \not= 0$. The function 
\begin{equation}
f(r_{12})=\frac{1}{2} r_{12}
\end{equation}
thus gives the behavior of the wave function at small interelectronic distances. Writing $r_{12}=||\b{r}_2 - \b{r}_1||=\sqrt{r_1^2 + r_2^2 -2 r_1 r_2 \cos \theta}$ where $\theta$ is the relative angle between the position vectors $\b{r}_1$ and $\b{r}_2$ of the two electrons, the function $f(r_{12})$ can be written as a partial-wave expansion
\begin{equation}
f(r_{12}) = \sum_{\ell = 0}^{\infty} f_\ell \; P_\ell (\cos \theta),
\label{fr12inf}
\end{equation}
where $P_\ell$ are the Legendre polynomials and the coefficients $f_\ell$ are
\begin{equation}
f_\ell = \frac{1}{2} \left( \frac{1}{2\ell + 3}\frac{r_<^{\ell+2}}{r_>^{\ell+1}} - \frac{1}{2\ell - 1}\frac{r_<^{\ell}}{r_>^{\ell-1}}  \right),
\end{equation}
with $r_<=\min(r_1,r_2)$ and $r_>=\max(r_1,r_2)$. The coefficients $f_\ell$ decrease slowly with $\ell$ when $r_1$ and $r_2$ are similar. In particular, for $r_1=r_2$, we have $f_\ell \sim \ell^{-2}$ as $\ell \to \infty$~\cite{FraMusLupTou-JJJ-XX-note1}. Therefore, the approximation of $f(r_{12})$ by a truncated partial-wave expansion, $\ell \leq L$,
\begin{equation}
f_L(r_{12}) = \sum_{\ell = 0}^{L} f_\ell \; P_\ell (\cos \theta),
\end{equation}
also converges slowly with $L$ near $r_{12}=0$. This is illustrated in Figure~\ref{cusp_plot} (left) which shows 1+$f_L(r_{12})$ as a function of $\theta$ for $r_1 = r_2 = 1$ bohr for increasing values of the maximal angular momentum $L$. Comparing with the converged value corresponding to $L\to \infty$ [Eq.~(\ref{fr12inf})], it is clear that the convergence near the singularity at $\theta=0$ is indeed painstakingly slow.

This slow convergence of the wave function near the electron-electron coalescence leads to the slow $L^{-4}$ power-law convergence of the partial-wave increments to the correlation energy~\cite{Sch-PR-62,Sch-INC-63,CarSilMet-JCP-79,KutMor-JCP-92,GriLud-JPB-02} or, equivalently, to the $L^{-3}$ power-law convergence of the truncation error in the correlation energy~\cite{Hil-JCP-85,God-SJMA-09}.

\begin{figure*}
\includegraphics[scale=0.30,angle=-90]{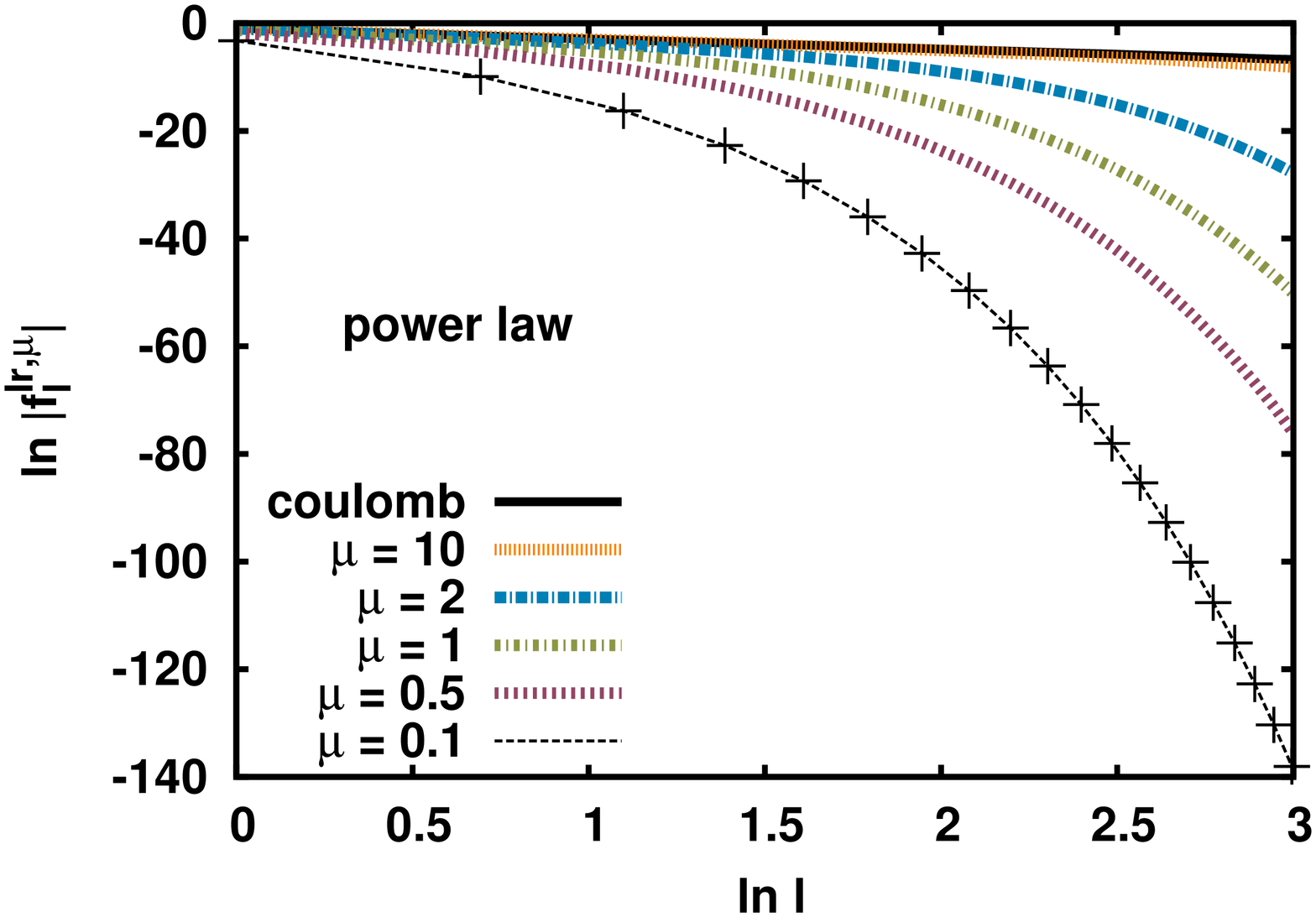}
\includegraphics[scale=0.30,angle=-90]{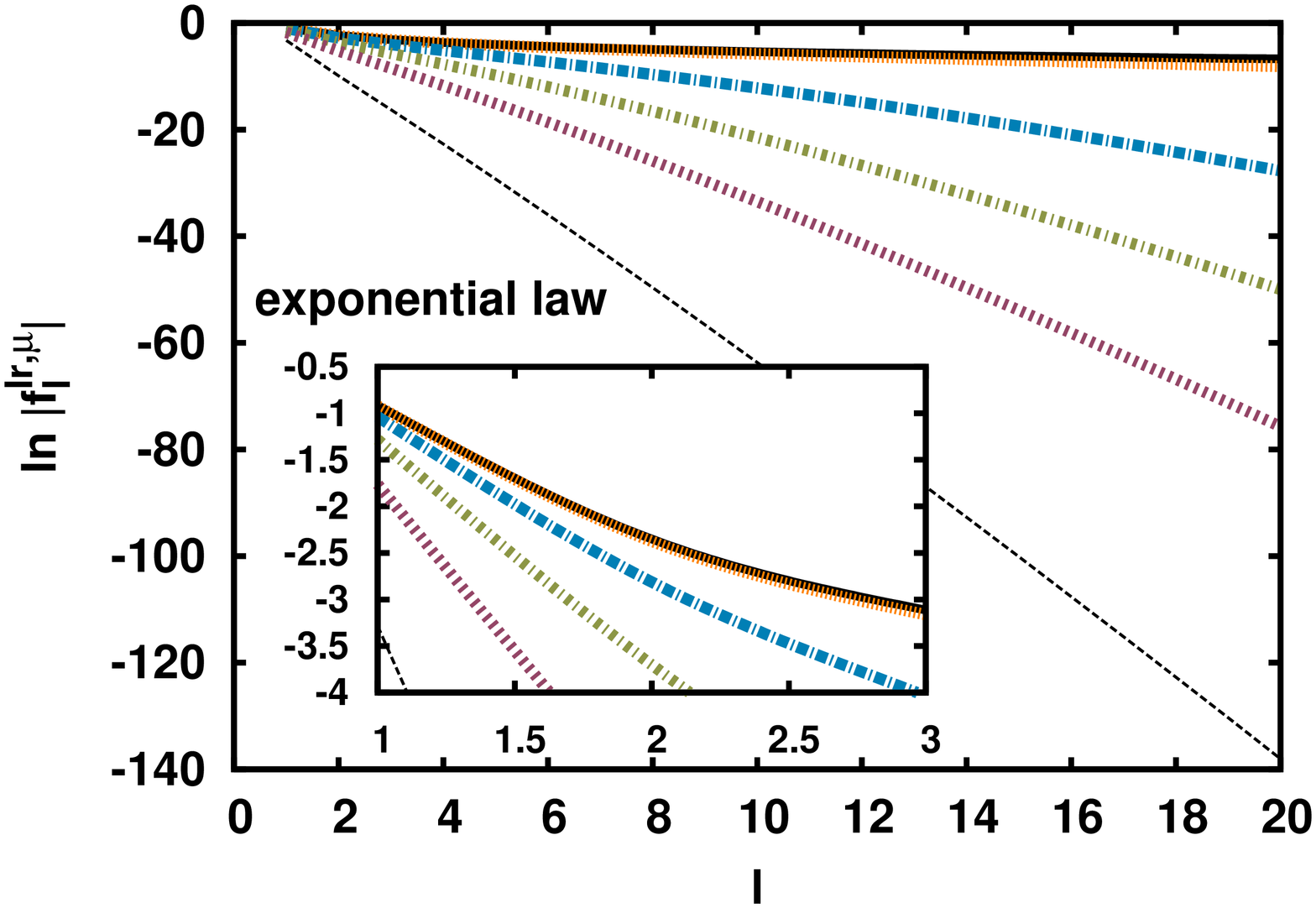}
\caption{Convergence rate of the coefficients $f\lr_\ell$ of the partial-wave expansion with respect to $\ell$ for $\ell \geq 1$, for several values of the range-separation parameter $\mu$ (in bohr$^{-1}$) and for the Coulomb case ($\mu\to\infty$). On the left: Plot of $\ln|f\lr_\ell|$ vs. $\ln \ell$ which is linear for a power-law convergence. On the right: Plot of $\ln|f\lr_\ell|$ vs. $\ell$ which is linear for an exponential-law convergence. The curves for the Coulomb interaction and for $\mu=10$ are nearly superimposed.}
\label{cvrate_plot}
\end{figure*}
     
\subsection{Long-range interaction}
\label{lr_case}

For systems with the long-range electron-electron interaction $w_{\text{ee}}\lr(r_{12})=\erf(\mu r_{12})/r_{12}$, the behavior of the wave function for small interelectronic distances $r_{12}$ was determined by Gori-Giorgi and Savin~\cite{GorSav-PRA-06}
\begin{equation}
\frac{\Psi\lr (r_{12})}{\Psi\lr(0)}  = 1 + r_{12} p_1(\mu r_{12}) + O(r_{12}^4),
\end{equation}
where the function $p_1(y)$ is given by
\begin{equation}
p_1(y) = \frac{e^{-y^2}-2}{2 \sqrt{\pi} y} + \left( \frac{1}{2} + \frac{1}{4 y^2} \right) \erf(y).
\end{equation}
We thus need to study the function 
\begin{equation}
f\lr(r_{12}) = r_{12} p_1(\mu r_{12}).
\end{equation}
For a fixed value of $\mu$, and for $r_{12} \ll 1/\mu$, it yields
\begin{equation}
f\lr(r_{12}) = \frac{\mu}{3\sqrt{\pi}} r_{12}^2 + O(r_{12}^4),
\label{fexpand}
\end{equation}
which exhibits no linear term in $r_{12}$, \ie no electron-electron cusp. On the other hand, for $\mu \to \infty$ and $r_{12} \gg 1/\mu$, we obtain
\begin{equation}
f^{\text{lr},\mu\to\infty}(r_{12}) = \frac{1}{2} r_{12} + O(r_{12}^2),
\end{equation}
\ie the Coulomb electron-electron cusp is recovered.
The function $f\lr(r_{12})$ thus makes the transition between the cuspless long-range wave function and the Coulomb wave function.

As for the Coulomb case, we write $f\lr(r_{12})$ as a partial-wave expansion
\begin{equation}
f\lr(r_{12}) = \sum_{\ell = 0}^{\infty} f\lr_\ell \; P_\ell (\cos \theta),
\end{equation}
and calculate with \textsc{Mathematica}~\cite{Math9-PROG-12} the coefficients $f\lr_\ell$ for each $\ell$
\begin{equation}
f\lr_\ell = \frac{2\ell+1}{2}\int_{-1}^{1} f\lr(r_{12}) P_\ell (x) \d x,
\end{equation}
with $x = \cos \theta$, $r_{12}=\sqrt{r_1^2 + r_2^2 -2 r_1 r_2 x}$, and using the following explicit expression for $P_\ell (x)$
\begin{equation}
P_\ell (x) = 2^\ell \sum_{k=0}^\ell {\ell \choose k} {\frac{\ell+k-1}{2} \choose \ell}  x^k.
\end{equation}
Since the partial-wave expansion of the first term in $r_{12}^2$ in Eq.~(\ref{fexpand}) terminates at $\ell=1$, we expect a fast convergence with $\ell$ of $f\lr_\ell$, for $\mu$ small enough, and thus also a fast convergence of the truncated partial-wave expansion
\begin{equation}
f\lr_L(r_{12}) = \sum_{\ell = 0}^{L} f\lr_\ell \; P_\ell (\cos \theta).
\end{equation}
Plots of this truncated partial-wave expansion for $\mu=0.5$ in Figure~\ref{cusp_plot} (right) confirm this expectation. The Coulomb singularity at $\theta=0$ has disappeared and the approximation $f\lr_L(r_{12})$ converges indeed very fast with $L$, being converged to better than $0.001$ a.u. already at $L=2$. 

We study in detail the dependence of the coefficients of the partial-wave expansion $f\lr_\ell$ on $\ell$. We compare two possible convergence behaviors, a power-law form
\begin{equation}
f\lr_\ell = A \; \ell^{-\alpha},
\label{poly}
\end{equation}
and an exponential-law form
\begin{equation}
f\lr_\ell = B \; \exp(-\beta \ell),
\label{exp}
\end{equation}
where $A$, $B$, $\alpha$ and $\beta$ are ($\mu$-dependent) parameters.

To determine which form best represents $f\lr_\ell$, in \Fig{cvrate_plot} we plot $\ln |f\lr_\ell|$ for $r_1=r_2=1$ as a function of $\ln \ell$ (left) and as a function of $\ell$ (right), for several values of $\mu$, as well as for the Coulomb case ($\mu \to \infty$)~\cite{FraMusLupTou-JJJ-XX-note4}. A straight line on the plot of $\ln |f\lr_\ell|$ vs. $\ln \ell$ indicates a power-law dependence, whereas a straight line on the plot of $\ln|f\lr_\ell|$ vs. $\ell$ indicates an exponential-law dependence.

For the Coulomb case (black curve nearly superimposed with the curve for $\mu=10$), we observe that the plot of $\ln|f_\ell|$ vs. $\ln \ell$ is linear, whereas the plot of $\ln|f_\ell|$ vs. $\ell$ is curved upward. This is expected for a power law $A \; \ell^{-\alpha}$ form. Moreover, we find $\alpha \approx 2$ as expected from Section~\ref{coulomb_case}. When going from large to small values of $\mu$, we observe that the plot of $\ln|f\lr_\ell|$ vs. $\ln \ell$ becomes more and more curved downward, and the plot of $\ln|f\lr_\ell|$ vs. $\ell$ becomes more and more linear. We thus go from a power-law dependence to an exponential-law dependence. Already for $\mu \leq 2$, the exponential law is a better description than the power law.

When $\mu$ decreases, the absolute value of the slope of the plot of $\ln|f\lr_\ell|$ vs. $\ell$ increases, \ie the convergence becomes increasingly fast. More precisely, we have found $\beta \approx 2.598 - 1.918 \ln \mu$ for $\mu \leq 2$.

The exponential convergence of the partial-wave expansion of the long-range wave function near the electron-electron coalescence implies a similar exponential convergence for the partial-wave expansion of the corresponding energy. The present study is thus consistent with the approximate exponential convergence of the partial-wave expansion of the energy of the helium atom in the presence of a long-range electron-electron interaction reported in Refs.~\onlinecite{SIRBU02,SIRBU03}. However, no quantitative comparison can be made between the latter work and the present work since the form of the long-range interaction is different.

\section{Convergence in one-electron atomic basis sets}
\label{sec:basisconv}

In this section, we study the convergence of the long-range wave function and correlation energy with respect to the size of the one-particle atomic basis. This problem is closely related to the convergence of the partial-wave expansion studied in the previous section. Indeed, for a two-electron atom in a singlet S state, it is possible to use the spherical-harmonic addition theorem to obtain the partial-wave expansion in terms of the relative angle $\theta$ between two electrons by products of the spherical harmonic part $Y_{\ell,m}$ of the one-particle atomic basis functions
\begin{equation}
P_\ell (\cos \theta) = \frac{4\pi}{2\ell+1} \sum_{m=-\ell}^\ell (-1)^{m} Y_{\ell,m}(\theta_1,\phi_1) Y_{\ell,-m}(\theta_2,\phi_2),
\end{equation}
where $\cos \theta = \cos \theta_1 \cos \theta_2 + \sin \theta_1 \sin \theta_2 \cos(\phi_1-\phi_2)$ with spherical angles $\theta_1$,$\phi_1$ and $\theta_2$,$\phi_2$. The partial-wave expansion can thus be obtained from a one-particle atomic basis, provided that the basis saturates the radial degree of freedom for each angular momentum $\ell$. In practice, of course, for the basis sets that we use, this latter condition is not satisfied. Nevertheless, one can expect the convergence with the maximal angular momentum $L$ of the basis to be similar to the convergence of the partial-wave expansion. 

For this study, we have analyzed the behavior of He, Ne, N$_2$, and H$_2$O at the same experimental geometries used in Ref.~\onlinecite{HALKIER98} ($R_{\rm N-N} = 1.0977$ \AA, $R_{\rm O-H} = 0.9572$ \AA\, and $\widehat{\rm HOH} = 104.52^{\circ}$). We performed all the calculations with the program \textsc{MOLPRO 2012}~\cite{Molproshort-PROG-12} using Dunning correlation-consistent cc-p(C)V$X$Z basis sets for which we studied the convergence with respect to the cardinal number $X$, corresponding to a maximal angular momentum of $L=X-1$ for He and $L=X$ for atoms from Li to Ne. We emphasize that the series of Dunning basis sets does not correspond to a partial-wave expansion but to a principal expansion~\cite{KloBakJorOlsHel-JPB-99,HelJorOls-BOOK-02} with maximal quantum number $N=X$ for He and $N=X+1$ for Li to Ne. The short-range exchange-correlation PBE density functional of Ref.~\onlinecite{GolWerStoLeiGorSav-CP-06} (which corresponds to a slight modification of the one of Ref.~\onlinecite{TouColSav-JCP-05}) was used in all range-separated calculations.

\begin{figure*}
\raisebox{-11em}{
\includegraphics{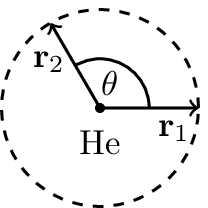}
}
\includegraphics[scale=0.27,angle=-90]{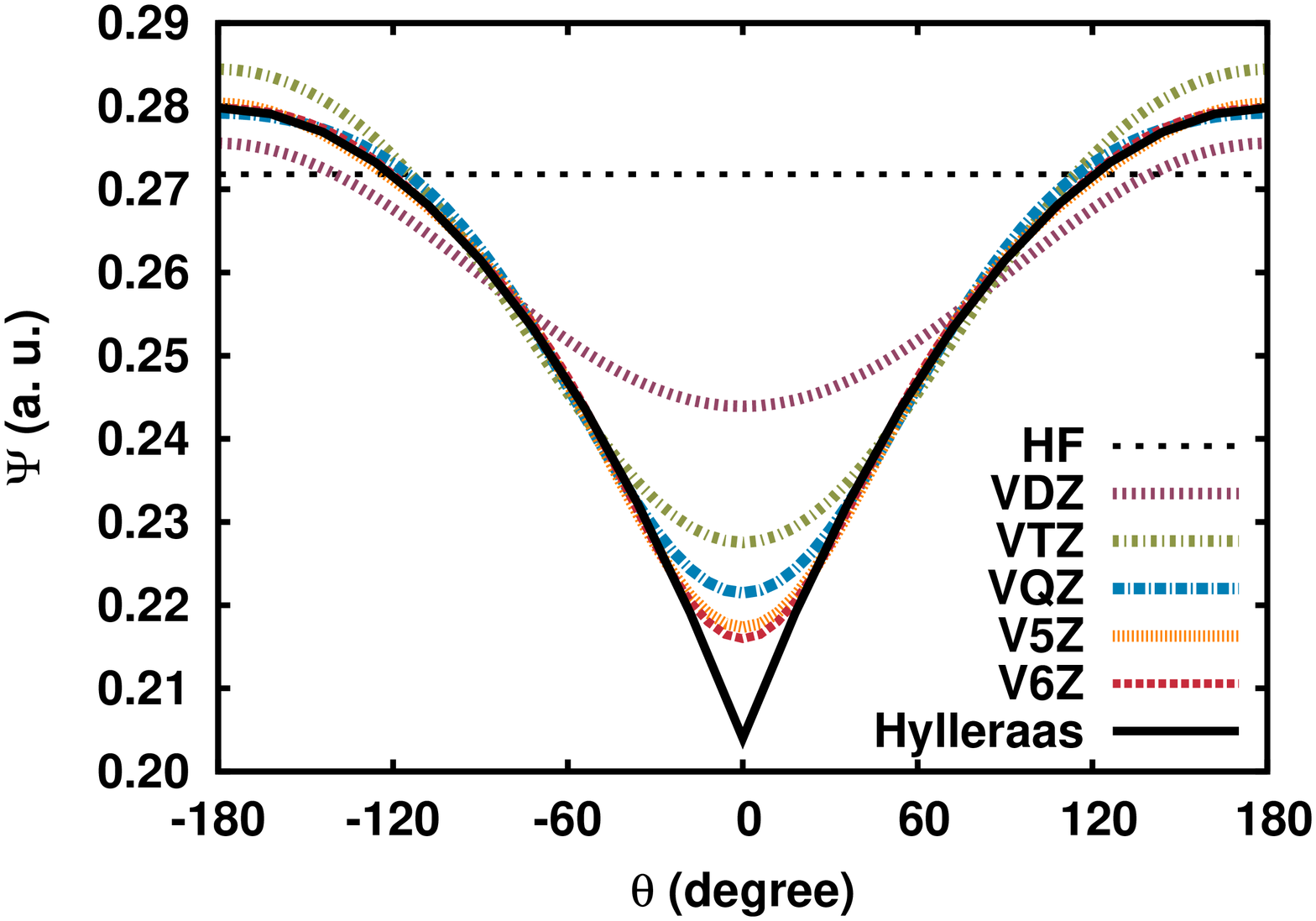}
\includegraphics[scale=0.27,angle=-90]{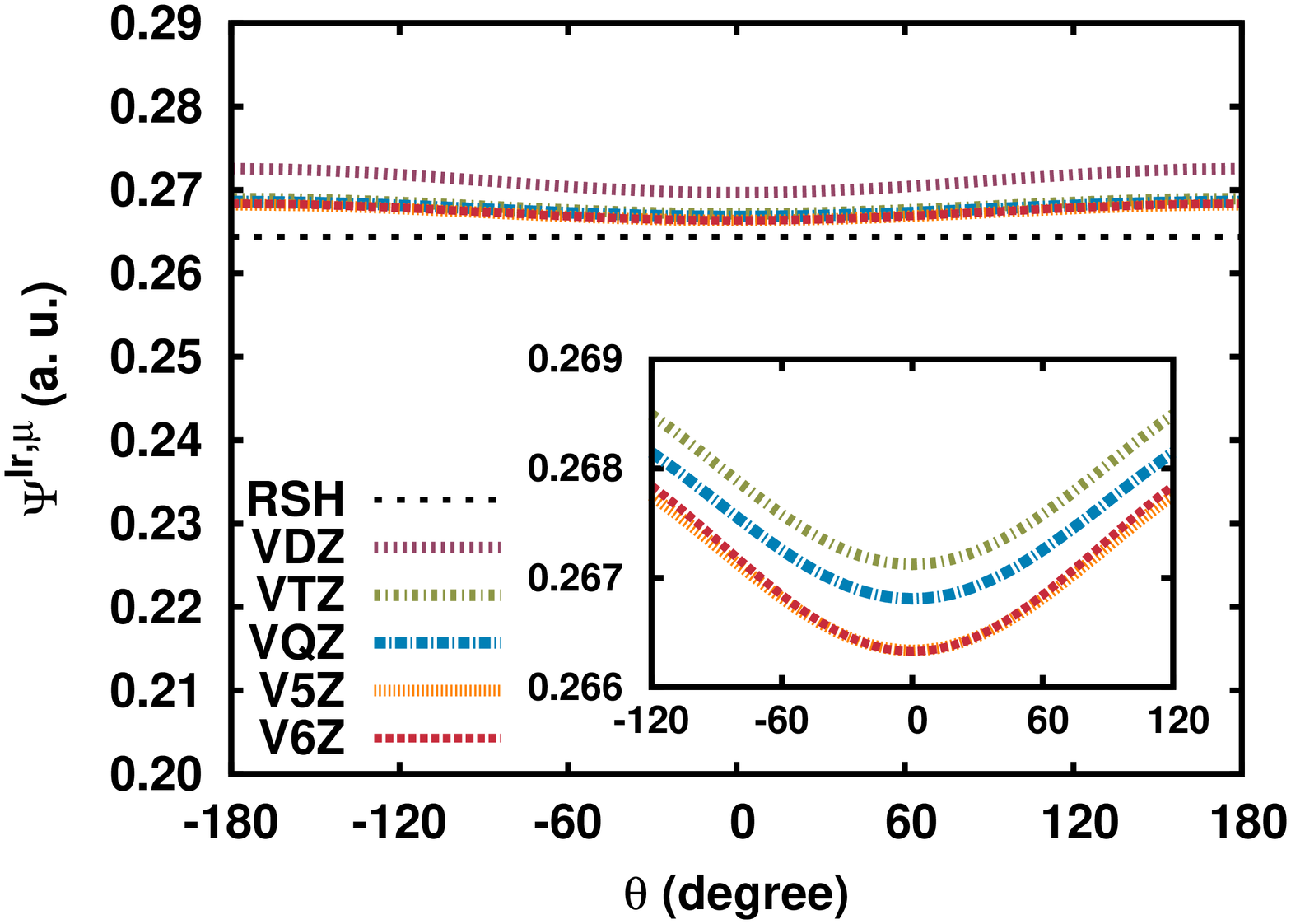}
\caption{FCI wave function of the He atom at the Cartesian electron coordinates $\b{r}_1=(0.5,0.,0.)$ bohr and $\b{r}_2=(0.5\cos \theta, 0.5 \sin \theta,0.)$ bohr, calculated with Dunning basis sets ranging from cc-pVDZ to cc-pV6Z (abbreviated as V$X$Z) and shown as a function of the relative angle $\theta$, for the standard Coulomb interaction case (left) and the long-range interaction case for $\mu=0.5$ bohr$^{-1}$ (right). For the case of the Coulomb interaction, an essentially exact curve has been calculated with a highly accurate 418-term Hylleraas-type wave function~\cite{FreHuxMor-PRA-84,BakFreHilMor-PRA-90,UmrGon-PRA-94}. For comparison, we also show the results obtained with the single-determinant HF and RSH wave functions (with the cc-pV6Z basis) which just give horizontal lines since they do not depend on $\theta$. In the insert plot on the right, the V5Z and V6Z curves are superimposed.
}
\label{cuspbasis}
\end{figure*}

\subsection{Convergence of the wave function}

We start by analyzing the convergence of the FCI ground-state wave function of the He atom with respect to the cardinal number $X$ of the cc-pV$X$Z basis sets. We perform a FCI calculation with the long-range Hamiltonian in Eq.~(\ref{HPsiEPsi}) using a fixed RSH density, calculated from the orbitals obtained in Eq.~(\ref{EminPhi}), in the short-range Hartree--exchange-correlation potential. To facilitate the extraction of the wave function from the program, we use the L\"owdin-Shull diagonal representation of the spatial part of the FCI wave function in terms of the spatial natural orbitals (NO) $\{\varphi_i^\mu\}$~\cite{LowShu-PR-56,SzaOst-BOOK-96}
\begin{equation}
\Psi\lr(\b{r}_1,\b{r}_2) = \sum_{i\geq 1} c_i^\mu \varphi_i^\mu(\b{r}_1) \varphi_i^\mu(\b{r}_2),
\end{equation}
where the coefficients $c_i^\mu$ are related to the NO occupation numbers $n_i^\mu$ by the relation $n_i^\mu = 2 |c_i^\mu|^2$. As the signs of $c_i^\mu$ are undetermined we have chosen a positive leading coefficient $c_1^\mu = \sqrt{n_1^\mu/2}$, and we assumed that all the other coefficients are negative $c_i^\mu = -\sqrt{n_i^\mu/2}$ for $i\geq 2$~\cite{GoeUmr-INC-00}. Even though it has been shown that, for the case of the Coulomb interaction, there are in fact positive coefficients in the expansion in addition to the leading one, for a weakly correlated system such as the He atom, these positive coefficients appear only in larger basis sets than the ones that we consider here and have negligible magnitude~\cite{SheMenGriBae-JCP-13,GieLee-JCP-13,GieLee-JCP-13b}.

In Figure~\ref{cuspbasis} we show the convergence of the FCI wave function $\Psi\lr(\b{r}_1,\b{r}_2)$ with the cardinal number $X$ for $\mu \to \infty$ which corresponds to the Coulomb interaction (left) and for $\mu = 0.5$ (right). The first electron is fixed at the Cartesian coordinates $\b{r}_1=(0.5,0.,0.)$ (measured from the nucleus) and the position of the second electron is varied on a circle at the same distance of the nucleus, $\b{r}_2=(0.5\cos \theta, 0.5 \sin \theta,0.)$. For the Coulomb interaction, we compare with the essentially exact curve obtained with a highly accurate 418-term Hylleraas-type wave function~\cite{FreHuxMor-PRA-84,BakFreHilMor-PRA-90,UmrGon-PRA-94}. The curve of $\Psi\lr(\b{r}_1,\b{r}_2)$ as a function of $\theta$ reveals the angular correlation between the electrons~\cite{FraMusLupTou-JJJ-XX-note2}. Clearly, correlation is much weaker for the long-range interaction. Note that a single-determinant wave function $\Phi(\b{r}_1,\b{r}_2)=\varphi_1(\b{r}_1) \varphi_1(\b{r}_2)$, where $\varphi_1$ is a spherically symmetric 1s orbital, does not depend on $\theta$, and the HF and RSH single-determinant wave functions indeed just give horizontal lines in Figure~\ref{cuspbasis}.

The fact that Figure~\ref{cuspbasis} resembles Figure~\ref{cusp_plot} confirms that the convergence with respect to $X$ is similar to the convergence of the partial-wave expansion with respect to $L$, and thus corroborates the relevance of the study of Section~\ref{sec:pwexpand} for practical calculations. As for the partial-wave expansion, the convergence with $X$ of the Coulomb wave function near the electron-electron cusp is exceedingly slow. The long-range wave function does not have an electron-electron cusp and converges much faster with $X$, the differences between the curves obtained with the cc-pV5Z and cc-pV6Z basis being smaller than 0.03 mhartree. Note, however, that the convergence of the long-range wave function seems a bit less systematic than the convergence of the Coulomb wave function, with the difference between the cc-pVQZ and cc-pV5Z basis being about 3 times larger than the difference between the cc-pVTZ and cc-pVQZ basis. This may hint to the fact the Dunning basis sets have been optimized for the Coulomb interaction and are not optimal for the long-range interaction. Finally, we note that we have found the same convergence behavior with the short-range exchange-correlation LDA density functional of Ref.~\onlinecite{PazMorGorBac-PRB-06}.

\subsection{Convergence of the correlation energy}\label{CVEc}

\begin{table*}
\caption{Valence MP2 correlation energies and their errors (in mhartree) for the Coulomb interaction ($E_\c$ and $\D E_\c$) and the long-range interaction at $\mu = 0.5$ bohr$^{-1}$ ($E_\c\lr$ and $\D E_\c\lr$) calculated with Dunning basis sets of increasing sizes for He, Ne, N$_2$ and H$_2$O. The errors are calculated with respect to the estimated CBS limit for the Coulomb interaction and with respect to the cc-pV6Z values for the long-range interaction.}
\label{ecMP2}
\begin{tabular}{l rrrrrrr rrrrrrr}
\hline\hline
& \multicolumn{11}{c}{Coulomb interaction} \\
& \multicolumn{2}{c}{He} && \multicolumn{2}{c}{Ne} && \multicolumn{2}{c}{N$_2$}  && \multicolumn{2}{c}{H$_2$O}\\
\cline{2-3} \cline{5-6}  \cline{8-9} \cline{11-12}
\multicolumn{1}{c}{Basis}
         & \multicolumn{1}{c}{$E_\c$} 
                        & \multicolumn{1}{c}{$\D E_\c$} 
                                   && \multicolumn{1}{c}{$E_\c$} 
                                                  & \multicolumn{1}{c}{$\D E_\c$} 
                                                              && \multicolumn{1}{c}{$E_\c$} 
                                                                             & \multicolumn{1}{c}{$\D E_\c$} 
                                                                                       && \multicolumn{1}{c}{$E_\c$} 
                                                                                                       & \multicolumn{1}{c}{$\D E_\c$} \\
\hline
cc-pVDZ  &  -25.828     &   11.549 && -185.523    &   134.577 && -306.297    & 114.903 &&  -201.621    & 98.479\\
cc-pVTZ  &  -33.138     &    4.239 && -264.323    &   55.777  && -373.683    &  47.517 &&  -261.462    & 38.638\\
cc-pVQZ  &  -35.478     &    1.899 && -293.573    &   26.527  && -398.749    &  22.451 &&  -282.798    & 17.302\\
cc-pV5Z  &  -36.407     &    0.970 && -306.166    &   13.934  && -409.115    &  12.085 &&  -291.507    &  8.593\\
cc-pV6Z  &  -36.807     &    0.570 && -311.790    &   8.310   && -413.823    &   7.377 &&  -295.205    &  4.895\\
CBS limit&  -37.377$^a$ &          && -320(1)$^b$ &           && -421(2)$^b$ &         &&  -300(1)$^b$ \\
\\
& \multicolumn{11}{c}{Long-range interaction} \\
& \multicolumn{2}{c}{He} && \multicolumn{2}{c}{Ne} && \multicolumn{2}{c}{N$_2$}  && \multicolumn{2}{c}{H$_2$O}\\
\cline{2-3} \cline{5-6}  \cline{8-9} \cline{11-12}
\multicolumn{1}{c}{Basis}
         & \multicolumn{1}{c}{$E_\c\lr$} 
                        & \multicolumn{1}{c}{$\D E_\c\lr$} 
                                   && \multicolumn{1}{c}{$E_\c\lr$} 
                                                  & \multicolumn{1}{c}{$\D E_\c\lr$} 
                                                              && \multicolumn{1}{c}{$E_\c\lr$} 
                                                                             & \multicolumn{1}{c}{$\D E_\c\lr$} 
                                                                                       && \multicolumn{1}{c}{$E_\c\lr$} 
                                                                                                       & \multicolumn{1}{c}{$\D E_\c\lr$} \\
\hline
cc-pVDZ & -0.131 &  0.227 &&     -0.692& 1.963&&   -20.178&       3.316&&     -6.462&        3.532 \\
cc-pVTZ & -0.262 &  0.096 &&     -1.776& 0.879&&   -22.663&       0.830&&     -8.956&        1.038 \\
cc-pVQZ & -0.322 &  0.036 &&     -2.327& 0.328&&   -23.263&       0.231&&     -9.626&        0.367\\
cc-pV5Z & -0.346 &  0.012 &&     -2.557& 0.098&&   -23.430&       0.064&&     -9.901&        0.092\\
cc-pV6Z & -0.358 &        &&     -2.655&      &&   -23.494&            &&     -9.993& \\
\hline\hline
\multicolumn{13}{l}{\footnotesize$^a$Taken from Ref.~\onlinecite{PatCenJezJezSza-JPCA-07} where it was obtained by a Gaussian-type geminal MP2 calculation.}\\
\multicolumn{13}{l}{\footnotesize$^b$Taken from Ref.~\onlinecite{HALKIER98} where it was estimated from R12-MP2 calculations.} 
\end{tabular}
\end{table*}

We also study the basis convergence of the long-range MP2 correlation energy, given in Eq.~(\ref{MP2orb}), calculated with RSH orbitals for He, Ne, N$_2$ and H$_2$O.

In Table~\ref{ecMP2} we show the valence MP2 correlation energies and their errors as a function of the cardinal number $X$ of the cc-pV$X$Z basis sets for X $\leq 6$. We compare the long-range MP2 correlation energies $E\lr_{\c,X}$ at $\mu = 0.5$ and the standard Coulomb MP2 correlation energies $E_{\c,X}$ corresponding to $\mu\to\infty$. For the case of the Coulomb interaction, the error is calculated as $\Delta E_{\c,X}= E_{\c,X} - E_{\c,\infty}$ where $E_{\c,\infty}$ is the MP2 correlation energy in the estimated CBS limit taken from Refs.~\onlinecite{PatCenJezJezSza-JPCA-07,HALKIER98}. For the range-separated case we do not have an independent estimate of the CBS limit of the long-range MP2 correlation energy for a given value of $\mu$. Observing that the difference between the long-range MP2 correlation energies for $X=5$ and $X=6$ is below 0.1 mhartree for $\mu=0.5$, we choose the cc-pV6Z result as a good estimate of the CBS limit. Of course, the accuracy of this CBS estimate will deteriorate for larger values of $\mu$, but in practice this is a good estimate for the range of values of $\mu$ in which we are interested, \ie $0 \leq \mu \leq 1$~\cite{FraMusLupTou-JJJ-XX-note3}. The error on the long-range correlation energy is thus calculated as $\D E_{\c,X}\lr = E_{\c, X}\lr - E_{\c, 6}\lr$. 

\begin{table*}
\caption{Results of the fits to the power and exponential laws of the Coulomb valence MP2 correlation energy error $\Delta E_{\c,X}$ and long-range valence MP2 correlation energy error $\Delta E\lr_{\c,X}$ for $\mu=0.5$ bohr$^{-1}$. Different ranges, $X_{\rm min} \leq X \leq X_{\rm max}$, for the cardinal number $X$ of the Dunning basis sets are tested. The parameters $A$ and $B$ are in mhartree. The squared Pearson correlation coefficients $r^2$ of the fits are indicated in \%. For each line, the largest value of $r^2$ is indicated in boldface.} 
\begin{tabular}{c cc cc rr rr rrr rr rr r}
\hline\hline
&&&&& \multicolumn{12}{c}{Coulomb interaction} \\
&& & && \multicolumn{5}{c}{Power law} &&& \multicolumn{5}{c}{Exponential law} \\
\cline{6-10} \cline{13-17}\\[-0.3cm]
      && \multicolumn{1}{c}{$X_{\rm min}$}
           &  \multicolumn{1}{c}{$X_{\rm max}$}
               &&  \multicolumn{1}{c}{$\alpha$}
                         &&  \multicolumn{1}{c}{$A$}
                                    &&  \multicolumn{1}{c}{$r^2$}
                                             &&&   \multicolumn{1}{c}{$\beta$}
                                                       &&  \multicolumn{1}{c}{$B$}
                                                                 &&  \multicolumn{1}{c}{$r^2$} \\ \hline
He    && 2 & 6 &&  2.749 && 81.84   && \textbf{99.82} &&& 0.749 && 44.04  && 98.55 \\
      && 3 & 6 &&  2.902 && 104.00  && \textbf{99.97} &&& 0.669 && 29.51  && 99.19 \\[0.1cm]
Ne    && 2 & 6 &&  2.543 && 843.36  && \textbf{99.59} &&& 0.696 && 479.76 && 99.00 \\
      && 3 & 6 &&  2.754 && 1169.90 && \textbf{99.93} &&& 0.636 && 355.28 && 99.37 \\[0.1cm]
N$_2$ && 2 & 6 &&  2.513 && 697.74  && \textbf{99.71} &&& 0.686 && 397.67 && 98.76 \\
      && 3 & 6 &&  2.693 && 923.76  && \textbf{99.98} &&& 0.621 && 286.90 && 99.16 \\[0.1cm]
H$_2$O&& 2 & 6 &&  2.742 && 717.04  && \textbf{99.52} &&& 0.751 && 391.27 && 99.09 \\
      && 3 & 6 &&  2.988 && 1051.28 && \textbf{99.92} &&& 0.690 && 288.61 && 99.39 \\
\\
&&&&& \multicolumn{12}{c}{Long-range interaction} \\
&& & && \multicolumn{5}{c}{Power law} &&& \multicolumn{5}{c}{Exponential law} \\
\cline{6-10} \cline{13-17}\\[-0.3cm]
      && \multicolumn{1}{c}{$X_{\rm min}$}
           &  \multicolumn{1}{c}{$X_{\rm max}$}
               &&  \multicolumn{1}{c}{$\alpha$}
                         &&  \multicolumn{1}{c}{$A$}
                                    &&  \multicolumn{1}{c}{$r^2$}
                                             &&&   \multicolumn{1}{c}{$\beta$}
                                                       &&  \multicolumn{1}{c}{$B$}
                                                                 &&  \multicolumn{1}{c}{$r^2$} \\ \hline
He    && 2 & 5 &&  3.128 && 2.35    && 96.71 &&& 0.974 && 1.68   && \textbf{99.77} \\
      && 3 & 5 &&  3.997 && 8.16    && 99.11 &&& 1.028 && 2.13   && \textbf{99.95} \\[0.1cm]
Ne    && 2 & 5 &&  3.189 && 22.06   && 95.04 &&& 0.998 && 15.97  && \textbf{99.18} \\
      && 3 & 5 &&  4.257 && 101.51  && 98.27 &&& 1.098 && 24.57  && \textbf{99.65} \\[0.1cm]
N$_2$ && 2 & 5 &&  4.257 && 73.26   && 98.63 &&& 1.313 && 44.49  && \textbf{99.97} \\
      && 3 & 5 &&  4.997 && 211.13  && 99.44 &&& 1.283 && 39.08  && \textbf{99.99} \\[0.1cm]
H$_2$O&& 2 & 5 &&  3.861 && 60.27   && 97.30 &&& 1.198 && 39.24  && \textbf{99.72} \\
      && 3 & 5 &&  4.686 && 196.20  && 97.63 &&& 1.210 && 41.50  && \textbf{99.33} \\
  \hline\hline
\end{tabular}
\label{eclrfit}
\end{table*}

The first observation to be made is that the long-range MP2 correlation energies only represent about 1 to 5 \% of the Coulomb MP2 correlation energies. Although the long-range correlation energy may appear small, it is nevertheless essential for the description of dispersion interactions for instance. The errors on the long-range MP2 correlation energies are also about two orders of magnitude smaller than the errors on the Coulomb MP2 correlation energies. 

Inspired by Ref.~\onlinecite{HALKIER99}, we compare two possible forms of convergence for the correlation energy: a power-law form
\begin{equation}\label{Ecpower}
E_{\text{c},X}\lr = E_{\c,\infty}\lr + A X^{-\alpha},
\end{equation}
and an exponential-law form
\begin{equation}\label{Ecexp}
E_{\text{c},X}\lr = E_{\c,\infty}\lr + B \exp(- \beta X),
\end{equation}
where $E_{\c,\infty}\lr$ is the CBS limit of the long-range correlation energy and $A$, $B$, $\alpha$, and $\beta$ are parameters depending on $\mu$, as in Section~\ref{sec:pwexpand}. In practice, we actually make linear fits of the logarithm of the error $\ln(\Delta E\lr_{\c, X})$ for the two forms:
\begin{equation}
\ln(\Delta E\lr_{\c,X}) = \ln A - \alpha \ln X,
\label{logexp}
\end{equation}
\begin{equation}
\ln(\Delta E\lr_{\c, X}) = \ln B - \beta X.
\label{logpower}
\end{equation}

In \Tab{eclrfit}, we show the results of the fits for the Coulomb interaction and the long-range interaction at $\mu=0.5$ using either the complete range of $X$ or excluding the value for $X=2$. We use the squared Pearson correlation coefficient $r^2$ as a measure of the quality of the fit. For the Coulomb interaction and for all the systems studied, the best fit is achieved with the power law $A X^{-\alpha}$ with $\alpha \approx 2.5 - 3$, which is roughly what is expected~\cite{HALKIER98}. We note however that the fits to the exponential law are also very good with $r^2>99 \%$ when the $X=2$ value is excluded. This explains why extrapolations of the total energy based on an exponential formula have also been used for the case of the Coulomb interaction~\cite{Fel-JCP-92,PetDun-JPC-95}. For the case of the long-range interaction, the difference between the fits to the power law and to the exponential law is much bigger. The best fit is by far obtained for the exponential law with $r^2>99 \%$ for all systems and ranges of $X$ considered. This exponential convergence of the long-range correlation energy with respect to $X$ is in accordance with the exponential convergence of the partial-wave expansion of the long-range wave-function observed in Section~\ref{sec:pwexpand}.

We have also performed fits for several other values of $\mu$ between $0.1$ and $1$ and always obtained an exponential convergence of the long-range valence MP2 correlation energy with respect to $X$. However, contrary to what was observed for the partial-wave expansion, we found that when $\mu$ decreases $\beta$ also decreases a bit for the four systems considered. In other words, when the interaction becomes more long range, the convergence of the long-range correlation energy becomes slower. This surprising result may be due to the fact that the cc-pV6Z result may not be as good an estimate of the CBS limit when $\mu$ increases. When $\mu$ decreases, the prefactor $B$ decreases and goes to zero for $\mu=0$, as expected. Moreover, we have also checked that we obtain very similar results for the long-range all-electron MP2 correlation energy (including core excitations) with cc-pCV$X$Z basis sets.

We note that Prendergast {\it et al.}~\cite{PreNolFilFahGre-JCP-01} have argued that the removal of the electron-electron cusp in a small region around the coalescence point does not significantly improve the convergence of the energy in the millihartree level of accuracy. At first sight, their conclusion might appear to be in contradiction with our observation of the exponential convergence of the long-range correlation energy with $X$. There are however important differences between the two studies: (1) their form of long-range interaction is different from ours, (2) they consider interelectronic distance ``cutoffs'' of $r_{c} \lesssim 0.8$ bohr whereas we consider larger ``cutoffs'' $r_{c}=1/\mu \geq 1$ bohr, (3) they do not investigate exponential-law versus power-law convergence.

Finally, in the Appendix we provide a complement analysis of the basis convergence of the correlation energy of the He atom for truncated configuration-interaction (CI) calculations in natural orbitals. The analysis shows that, contrary to the case of the Coulomb interaction, the convergence of the long-range correlation energy is no longer limited by the truncation rank the CI wave function but by the basis convergence of the natural orbitals themselves. This result is consistent with an exponential basis convergence of the long-range correlation energy.

\subsection{Extrapolation scheme}

For the long-range interaction case, since both the RSH energy and the long-range correlation energy have an exponential convergence with respect to the cardinal number $X$, we propose to extrapolate the total energy to the CBS limit by using a three-point extrapolation scheme based on an exponential formula. Suppose that we have calculated three total energies $E_X$, $E_Y$, $E_Z$ for three consecutive cardinal numbers $X$, $Y=X+1$, $Z=X+2$. If we write
\begin{equation}
E_X = E_{\infty} + B \exp(-\beta X),
\label{eX}
\end{equation}
\begin{equation}
E_Y = E_{\infty} + B \exp(-\beta Y),
\label{eY}
\end{equation}
\begin{equation}
E_Z = E_{\infty} + B \exp(-\beta Z), 
\label{eZ}
\end{equation}
and eliminate the unknown parameters $B$ and $\beta$, we obtain the following estimate of the CBS-limit total energy $E_{\infty}$
\begin{equation}
E_{\infty} = E_{XYZ} = \frac{E_Y^2 - E_X E_Z}{2E_Y - E_X - E_Z}.
\label{EXYZ}
\end{equation}

In \Tab{3ptsext}, we report the errors on the RSH+lrMP2 total energy, $E^\mu = E_\RSH^\mu + E_\text{c,MP2}\lr$, obtained with the three-point extrapolation formula using either $X = 2$, $Y = 3$, $Z=4$ ($\Delta E_{\rm DTQ}^\mu$) or $X = 3$, $Y = 4$, $Z=5$ ($\Delta E_{\rm TQ5}^\mu$), and we compare with the errors obtained with each cc-pV$X$Z basis set from $X=2$ to $X=5$ ($\Delta E_X^\mu$). Here again the errors are calculated with respect to the cc-pV6Z total energy, for several values of the range-separation parameter $\mu=0.1$, $0.5$, $1.0$, and only valence excitations are included in the MP2 calculations. For all the systems studied the errors $\Delta E_{\rm DTQ}^\mu$ are less than 1.5 mhartree. For Ne, N$_2$, and H$_2$O, these $\Delta E_{\rm DTQ}^\mu$ errors are significantly smaller (by a factor of about 3 to 15) than the errors $\Delta E_{\rm Q}^\mu$ obtained with the largest basis used for the extrapolation, and are overall comparable with the errors $\Delta E_{\rm 5}^\mu$. Thus, the three-point extrapolation formula with $X = 2$, $Y = 3$, $Z=4$ provides a useful CBS extrapolation scheme for range-separated DFT. Except for He, the errors $\Delta E_{\rm TQ5}^\mu$ are negative (\ie, the extrapolation overshoots the CBS limit) and larger than the errors $\Delta E_{\rm 5}^\mu$. Thus, the three-point extrapolation scheme with $X = 3$, $Y = 4$, $Z=5$ does not seem to be useful.

These conclusions extend to calculations including core excitations with cc-pCV$X$Z basis sets, which are presented in \Tab{3ptsextcore}. All the errors are smaller than for the valence-only calculations. The errors $\Delta E_{\rm DTQ}^\mu$ are now less than 0.9 mhartree, and are smaller or comparable to the errors $\Delta E_{\rm 5}^\mu$. The errors $\Delta E_{\rm TQ5}^\mu$ are always negative and are overall larger than the errors $\Delta E_{\rm DTQ}^\mu$.

\begin{table}
\caption{Errors (in mhartree) on the total RSH+lrMP2 energy, $E^\mu = E_\RSH^\mu + E_\text{c,MP2}\lr$, obtained with cc-pV$X$Z basis sets from $X=2$ to $X=5$ ($\Delta E_X^\mu = E_X^\mu - E_6^\mu$) and with the three-point extrapolation formula of Eq.~(\ref{EXYZ}) using $X = 2$, $Y = 3$, $Z=4$ ($\Delta E_{\rm DTQ}^\mu = E_{\rm DTQ}^\mu - E_6^\mu$) or $X = 3$, $Y = 4$, $Z=5$ ($\Delta E_{\rm TQ5}^\mu = E_{\rm TQ5}^\mu - E_6^\mu$). The errors are calculated with respect to the cc-pV6Z total energy for several values of the range-separation parameter $\mu$ (in bohr$^{-1}$). Only valence excitations are included in the MP2 calculations.}
\label{3ptsext}
\begin{tabular}{llrrrrrr}
\hline\hline
\phantom{xxxxx} & $\mu$ \phantom{xx}
         & \multicolumn{1}{c}{$\Delta E_{\rm D}^\mu$}
                 &  \multicolumn{1}{c}{$\Delta E_{\rm T}^\mu$}
                         &\multicolumn{1}{c}{$\Delta E_{\rm Q}^\mu$}
                                 & \multicolumn{1}{c}{$\Delta E_5^\mu$}
                                         & \multicolumn{1}{c}{$\Delta E_{\rm DTQ}^\mu$}
                                                 & \multicolumn{1}{c}{$\Delta E_{\rm TQ5}^\mu$}\\ 
\hline
He & 0.1 &    8.508& 0.772& 0.261& 0.089& 0.224 & 0.003\\
   & 0.5 &    8.488& 0.781& 0.245& 0.078& 0.205 & 0.002\\
   & 1.0 &    8.258& 0.924& 0.259& 0.078& 0.192 & 0.011\\[0.1cm]
Ne & 0.1 &	72.999&	20.215&	5.842&	0.716& 0.464 & -2.127\\
   & 0.5 &	74.523&	20.337&	5.763&	0.751& 0.401 & -1.876\\
   & 1.0 &	79.311&	20.962&	5.726&	0.803& 0.342 & -1.548\\[0.1cm]
N$_2$ & 0.1 &	47.061&	13.026&	4.136&	0.853& 0.993 & -1.069\\
      & 0.5 &	51.581&	13.406&	4.090&	0.810& 1.083 & -0.972\\
      & 1.0 &	61.053&	15.108&	4.513&	0.868& 1.337 & -1.043\\[0.1cm]
H$_2$O& 0.1 &	54.861 &	15.229 &	5.005 &	0.857& 1.451 & -1.975\\
      & 0.5 &	55.850 &	14.736 &	4.499 &	0.726& 1.105 & -1.475\\
       & 1.0 &	61.013 &	15.212 &	4.423 &	0.724& 1.099 & -1.206\\
\hline\hline
\end{tabular}
\end{table}

\begin{table}
\caption{Errors (in mhartree) on the total RSH+lrMP2 energy, $E^\mu = E_\RSH^\mu + E_\text{c,MP2}\lr$, obtained with cc-pCV$X$Z basis sets from $X=2$ to $X=5$ ($\Delta E_X^\mu = E_X^\mu - E_6^\mu$) and with the three-point extrapolation formula of Eq.~(\ref{EXYZ}) using $X = 2$, $Y = 3$, $Z=4$ ($\Delta E_{\rm DTQ}^\mu = E_{\rm DTQ}^\mu - E_6^\mu$) or $X = 3$, $Y = 4$, $Z=5$ ($\Delta E_{\rm TQ5}^\mu = E_{\rm TQ5}^\mu - E_6^\mu$). The errors are calculated with respect to the cc-pCV6Z total energy for several values of the range-separation parameter $\mu$ (in bohr$^{-1}$). Core excitations are included in the MP2 calculations.}
\label{3ptsextcore}
\begin{tabular}{llrrrrrr}
\hline\hline
\phantom{xxxxx} & $\mu$ \phantom{xx}
         & \multicolumn{1}{c}{$\Delta E_{\rm D}^\mu$}
                 &  \multicolumn{1}{c}{$\Delta E_{\rm T}^\mu$}
                         &\multicolumn{1}{c}{$\Delta E_{\rm Q}^\mu$}
                                 & \multicolumn{1}{c}{$\Delta E_5^\mu$}
                                         & \multicolumn{1}{c}{$\Delta E_{\rm DTQ}^\mu$}
                                                 & \multicolumn{1}{c}{$\Delta E_{\rm TQ5}^\mu$}\\ 
\hline
Ne    & 0.1 & 70.932 & 18.941 & 4.929 & 0.522 & -0.240 & -1.501\\
      & 0.5 & 72.501 & 18.990 & 4.831 & 0.537 & -0.263 & -1.333\\
      & 1.0 & 77.497 & 19.517 & 4.775 & 0.554 & -0.250 & -1.140 \\[0.1cm]
N$_2$ & 0.1 & 43.528 & 10.237 & 2.334 & 0.459 & -0.126 & -0.125\\
      & 0.5 & 48.079 & 10.451 & 2.285 & 0.413 & 0.021 & -0.144\\
      & 1.0 & 57.942 & 12.118 & 2.677 & 0.467 & 0.227 & -0.209\\[0.1cm]
H$_2$O& 0.1 & 52.875 & 13.897 & 4.132 & 0.680 & 0.868 & -1.208\\
      & 0.5 & 53.936 & 13.350 & 3.614 & 0.534 & 0.541 & -0.891\\
      & 1.0 & 59.290 & 13.789 & 3.527 & 0.521 & 0.539 & -0.724\\
\hline\hline
\end{tabular}	
\end{table}

We have also tested a more flexible extrapolation scheme where the RSH energy and the long-range MP2 correlation energy are exponentially extrapolated independently but we have not found significant differences. On the contrary, one may want to use a less flexible two-point extrapolation formula using a predetermined value for $\beta$. The difficulty with such an approach is to choose the value of $\beta$, which in principle should depend on the system, on the range-separated parameter $\mu$, and on the long-range wave-function method used. For this reason, we do not consider two-point extrapolation schemes.

\section{Conclusions}

We have studied in detail the basis convergence of range-separated DFT. We have shown that the partial-wave expansion of the long-range wave function near the electron-electron coalescence converges exponentially with the maximal angular momentum $L$. We have also demonstrated on four systems (He, Ne, N$_2$, and H$_2$O) that the long-range MP2 correlation energy converges exponentially with the cardinal number $X$ of the Dunning basis sets cc-p(C)V$X$Z. This contrasts with the slow $X^{-3}$ convergence of the correlation energy for the standard case of the Coulomb interaction. Due to this exponential convergence, the extrapolation to the CBS limit is less necessary for range-separated DFT than for standard correlated wave function methods. Nevertheless, we have proposed a CBS extrapolation scheme for the total energy in range-separated DFT based on an exponential formula using calculations from three cardinal numbers $X$. For the systems studied, the extrapolation using $X=2,3,4$ gives an error on the total energy with respect to the estimated CBS limit which is always smaller than the error obtained with a single calculation at $X=4$, and which is often comparable or smaller than the error obtained with a calculation at $X=5$. 

We expect the same convergence behavior for range-separated DFT methods in which the long-range part is treated by configuration interaction, coupled-cluster theory, random-phase approximations, or density-matrix functional theory. Finally, it should be pointed out that this rapid convergence is obtained in spite of the fact that the Dunning basis sets have been optimized for the case of the standard Coulomb interaction. The construction of basis sets specially optimized for the case of the long-range interaction may give yet a faster and more systematic convergence.

\label{app:rank}

\begin{table}[t]
\caption{Truncated CI correlation energies (in mhartree) of the He atom for the Coulomb interaction and for the long-range interaction (at $\mu=0.5$ bohr$^{-1}$) for different Dunning basis sets cc-pV$X$Z (abbreviated as V$X$Z) and truncation ranks.}
\begin{tabular}{ll rr rr rr rr r}
\hline\hline
	&& \multicolumn{9}{c}{Coulomb interaction} \\
rank	&& 	VDZ && 	VTZ && VQZ && V5Z && V6Z \\ \hline
1s2s		&&-14.997	&&	-15.806	&&-16.087	&&-16.204	&&-16.212	\\
1s2s2p		&&-32.434	&&	-35.256	&&-35.664	&&-35.794	&&-35.808	\\
1s2s2p3s	&&		&&	-35.909	&&-36.425	&&-36.595	&&-36.618	\\
1s2s2p3s3p	&&		&&	-37.448	&&-38.068	&&-38.261	&&-38.291	\\
1s2s2p3s3p3d	&& 		&&	-39.079	&&-39.807	&&-39.999	&&-40.028	\\ \\
	&& \multicolumn{9}{c}{Long-range interaction} \\
rank	&& 	VDZ && 	VTZ && VQZ && V5Z && V6Z \\ \hline
1s2s	&&-0.018	&&-0.036	&&-0.038	&&-0.039	&&-0.039\\
1s2s2p	&&-0.155	&&-0.329	&&-0.415	&&-0.449	&&-0.468\\
1s2s2p3s	&&	&&-0.329	&&-0.415	&&-0.450	&&-0.468\\
1s2s2p3s3p	&&	&&-0.329	&&-0.415	&&-0.450	&&-0.469\\
1s2s2p3s3p3d	&&	&&-0.329	&&-0.416	&&-0.451	&&-0.470\\
\hline \hline
\end{tabular}
\label{rangcorrE}
\end{table}

\begin{table}[t]
\caption{Results of the fits to the power and exponential laws of the truncated CI correlation energy error for the Coulomb interaction, $\Delta E_{\rm c,X} = E_{\rm c,X} - E_{\rm c,6}$, and the long-range interaction, $\Delta E\lr_{\rm c,X} = E\lr_{\rm c,X} - E\lr_{\rm c,6}$ for $\mu = 0.5$ bohr$^{-1}$. The range for the cardinal number $X$ of the Dunning basis sets is $2 \leq X \leq 6$ for the 1s2s and 1s2s2p ranks, and $3 \leq X \leq 6$ for all the larger ranks. The parameters $A$ and $B$ are in mhartree. The squared Pearson correlation coefficients $r^2$ of the fits are indicated in $\%$. For each line, the largest value of $r^2$ is indicated in boldface.
}
\begin{tabular}{l rr rr rrr rr r r}
\hline\hline
&& \multicolumn{10}{c}{Coulomb interaction} \\
 && \multicolumn{4}{c}{Power law} &&& \multicolumn{4}{c}{Exponential law}\\
\cline{9-12} \cline{3-7}\\[-0.3cm]
rank                    && $\alpha$ & $A$ && $r^2$ &&&  $\beta$  & $B$ &&  $r^2$ \\
\hline
1s2s			&&5.081&	65.641	&& 	87.44&&&	1.618&	43.148 &&	\textbf{94.39}\\
1s2s2p			&&5.765&	241.923	&&	95.45&&&	1.797&	131.516&&	\textbf{98.80}\\
1s2s2p3s		&&6.616&	1219.794&&	95.63&&&	1.716&	140.022&&	\textbf{98.11}\\
1s2s2p3s3p		&&6.440&	1168.927&&	96.46&&&	1.668&	140.605&&	\textbf{98.65}\\
1s2s2p3s3p3d	        &&6.771&	1874.080&&	97.16&&&	1.751&	200.085&&	\textbf{99.07}\\ \\
&& \multicolumn{10}{c}{Long-range interaction} \\
 && \multicolumn{4}{c}{Power law} &&& \multicolumn{4}{c}{Exponential law} \\
\cline{9-12} \cline{3-7} \\[-0.3cm]
rank                    && $\alpha$ & $A$ && $r^2$ &&&  $\beta$  & $B$ &&  $r^2$ \\
\hline
1s2s			&&3.974&	0.321 &&	\textbf{99.45}&&&	1.206&	0.188&&	97.51\\
1s2s2p			&&3.058&	3.108 &&	96.28&&&	0.953&	2.249&&	\textbf{99.64}\\
1s2s2p3s		&&3.956&	11.283&&	99.01&&&	1.018&	2.991&&	\textbf{99.93}\\
1s2s2p3s3p		&&3.927&	10.979&&	99.02&&&	1.010&	2.938&&	\textbf{99.93}\\
1s2s2p3s3p3d	        &&3.892&	10.701&&	99.00&&&	1.001&	2.898&&	\textbf{99.93}\\
\hline\hline
\end{tabular}
\label{rankfit}
\end{table}

\section*{Acknowledgements}
We thank Andreas Savin, Trygve Helgaker, \'Eric Canc\`es, and Gabriel Stoltz for insightful discussions. This work was supported by French state funds managed by CALSIMLAB and the ANR within the Investissements d'Avenir program under reference ANR-11-IDEX-0004-02.

\vspace{0.5cm}
\appendix
\section{Convergence of the correlation energy for truncated CI calculations}

In this Appendix, we explore the basis convergence of the correlation energy of the He atom for truncated CI calculations for both the Coulomb and long-range interactions. For a given basis set and interaction, we start by performing a FCI calculation and generating the corresponding natural orbitals. We then use these natural orbitals in truncated CI calculations for increasing orbital active spaces 1s2s, 1s2s2p, 1s2s2p3s, 1s2s2p3s3p, and 1s2s2p3s3p3d. Table~\ref{rangcorrE} shows the Coulomb and long-range correlation energies for the different basis sets and truncation ranks. For the Coulomb interaction, the correlation energy for a fixed rank converges rapidly with the basis size, while the convergence with respect to the rank is much slower. For the long-range interaction, the correlation energy jumps by one order of magnitude when including the 2p natural orbital, which is consistent with the fact that the long-range interaction brings in first angular correlation effects~\cite{PolColLeiStoWerSav-IJQC-03}. The long-range correlation energy is essentially converged at rank 1s2s2p, and the overall convergence is now determined by the basis convergence of the natural orbitals.

Finally, we compare two possible forms for the convergence of the truncated CI correlation energies with the cardinal number $X$, the power law~\Eq{Ecpower} and the exponential law~\Eq{Ecexp}. Using as reference the results obtained with the cc-pV6Z basis set, we have calculated for the different truncation ranks the correlation energy errors for the Coulomb interaction, $\Delta E_{\rm c,X} = E_{\rm c,X} - E_{\rm c,6}$, and for the long-range interaction, $\Delta E\lr_{\rm c,X} = E\lr_{\rm c,X} - E\lr_{\rm c,6}$, and performed logarithmic fits as in Section~\ref{CVEc}. Table~\ref{rankfit} shows the results of the fits. For both the Coulomb and long-range interactions, for the rank 1s2s2p and larger, the best fit is achieved with the exponential law $B \exp(-\beta X)$. Thus, in comparison with the Coulomb interaction, the long-range interaction does not significantly change the basis convergence of the correlation energy at a fixed truncation rank. However, for the long-range interaction, this exponential convergence at a fixed truncation rank becomes the dominant limitation to the overall basis convergence.


\begin{thebibliography}{10}
\bibitem{FraMusLupTou-JJJ-XX-note3}
For $\mu=1$ the difference between the long-range MP2 correlation energies for
  $X=5$ and $X=6$ is still as small as about 0.2 mhartree.
\end{thebibliography}

\end{document}